\documentclass[12pt]{article}

\newenvironment{Itemize}{\begin{list}{$\bullet$}%
{\setlength{\topsep}{0.2mm}\setlength{\partopsep}{0.2mm}%
\setlength{\itemsep}{0.2mm}\setlength{\parsep}{0.2mm}}}%
{\end{list}}
\newcounter{enumct}
\newenvironment{Enumerate}{\begin{list}{\arabic{enumct}.}%
{\usecounter{enumct}\setlength{\topsep}{0.2mm}%
\setlength{\partopsep}{0.2mm}\setlength{\itemsep}{0.2mm}%
\setlength{\parsep}{0.2mm}}}{\end{list}}

\newcommand{\lessim}{\raisebox{-0.8mm}%
{\hspace{1mm}$\stackrel{<}{\sim}$\hspace{1mm}}}
 
\newlength{\abstwidth}
\begin{document}
 
%set sloppy attitude to line breaks
\sloppy

%do not number first page
\pagestyle{empty}
 
\begin{flushright}
LU TP 99--42\\
hep-ph/0001032\\
December 1999
\end{flushright}
 
\vspace{\fill}
\begin{center}
{\LARGE\bf Recent Progress in PYTHIA\footnote{To appear in the
Proceedings of the QCD and Weak Boson Physics workshop
in preparation for Run II at the Fermilab Tevatron}}\\[10mm]
{\Large Torbj\"orn Sj\"ostrand\footnote{torbjorn@thep.lu.se}} \\[3mm]
{\it Department of Theoretical Physics,}\\[1mm]
{\it Lund University, Lund, Sweden}
\end{center}
 
\vspace{\fill}
 
\begin{center}
{\bf Abstract}\\[2ex]
\begin{minipage}{\abstwidth}
PYTHIA is a general-purpose event generator for multiparticle production
in high-energy physics. After a general introduction and program news 
survey, some areas of recent physics progress are considered: 
the matching to matrix elements, especially for $W/Z$ production;
charm and bottom hadronization; multiple interactions; and
interconnection effects. The report concludes with some words on the 
future, specifically the ongoing transition to C++.
\end{minipage}
\end{center}
 
\vspace{\fill}
 
\clearpage
\pagestyle{plain}
\setcounter{page}{1}

\section{Introduction}

A general-purpose generator in high-energy physics should address a 
number of physics aspects, such as:
\begin{Itemize}
\item the matrix elements for a multitude of hard subprocesses of interest,
\item the convolution with parton distributions to obtain the
hard-scattering kinematics and cross sections,
\item resonance decays that (more or less) form part of the hard subprocess
(such as $W$, $Z$, $t$ or $h$),
\item initial- and final-state QCD and QED showers (or, as an alternative,
higher-order matrix elements, including a consistent treatment of 
vir\-tual-correction terms), 
\item multiple parton--parton interactions,
\item beam remnants,
\item hadronization,
\item decay chains of unstable particles, and
\item general utility and analysis routines (such as jet finding).
\end{Itemize}
Furthermore, one must be prepared for unexpected or less 
conventional effects, that could modify the assumed behaviour:
the strong-interaction dynamics in QCD remains unsolved and
thereby unpredictable in an absolute sense.
 
The PYTHIA 6.1 program was released in March 1997, as a merger of 
JETSET 7.4, PYTHIA 5.7 \cite{pythia} and SPYTHIA \cite{spythia}.
It covers all of the above areas. The current subversion is
PYTHIA 6.136, which contains over 50,000 lines of Fortran 77 code.
The code, manuals and sample main programs may be found at\\
\texttt{http://www.thep.lu.se/}$\sim$\texttt{torbjorn/Pythia.html}~.

The two other programs of a similar scope are HERWIG \cite{herwig}\\
\texttt{http://hepwww.rl.ac.uk/theory/seymour/herwig/} \\
and ISAJET \cite{isajet}\\
\texttt{ftp://penguin.phy.bnl.gov/pub/isajet}~.\\
For parton-level processes, many more programs have been written.
The availability of several generators provides for useful cross-checks 
and a healthy competition. Since the physics of a complete hadronic 
event is very complex and only partially understood from first principles,
one should not prematurely converge on one single approach.
 
\section{PYTHIA 6.1 Main News}

Relative to previous versions, main news in PYTHIA 6.1 include
\begin{Itemize}
\item a renaming of the old JETSET program elements to begin with 
\texttt{PY}, therefore now standard throughout,
\item new SUSY processes and improved SUSY simulation relative to 
SPYTHIA, and new PDG codes for sparticles,
\item new processes for Higgs (including doubly-char\-ged in 
left--right symmetric models), technicolor, \ldots, 
\item several improved resonance decays, including an alternative 
Higgs mass shape,
\item some newer parton distributions, such as CTEQ5 \cite{cteq5}
\item initial-state showers matched to some matrix elements,
\item new options for final-state gluon splitting to a pair of $c/b$ 
quarks and modified modelling of initial-state flavour excitation,
\item an energy-dependent $p_{\perp\mathrm{min}}$ in multiple interactions, 
\item an improved modelling of the hadronization of small-mass strings,
of importance especially for $c/b$, and
\item a built-in package for one-dimensional histo\-grams (based on GBOOK). 
\end{Itemize}
Some of these topics will be further studied below. Other improvements, 
of less relevance for $\overline{p}p$ colliders, include
\begin{Itemize}
\item improved modelling of gluon emission off $c/b$ quarks in $e^+e^-$,
\item colour rearrangement options for $W^+W^-$ events, 
\item a Bose-Einstein algorithm expanded with new options,
\item a new alternative baryon production scheme \cite{patrik},
\item QED radiation off an incoming muon,
\item a new machinery to handle real and virtual photon 
fluxes, cross sections and parton distributions \cite{christer}, and
\item new standard interfaces for the matching to external generators of
two, four and six fermions (and of two quarks plus two gluons) in 
$e^+e^-$.
\end{Itemize}

The current list of over 200 different subprocesses covers topics 
such as hard and soft QCD, heavy flavours, DIS and $\gamma\gamma$, 
electroweak production of $\gamma^*/Z^0$ and $W^{\pm}$ (singly or in 
pairs), production of a light or a heavy Standard Model Higgs, or of 
various Higgs states in supersymmetric (SUSY) or left--right symmetric 
models, SUSY particle production (sfermions, gauginos, etc.), 
technicolor, new gauge bosons, compositeness, and leptoquarks. 

Needless to say, most users will still find that their particular area
of interest is not as well addressed as could be wished. In some areas,
progress will require new ideas, while lack of time is the limiting 
factor in others.

\section{Matching to  Matrix Elements}

The matrix-element (ME) and parton-shower (PS) approaches to higher-order
QCD corrections both have their advantages and disadvantages. The former
offers a systematic expansion in orders of $\alpha_s$, and a powerful
machinery to handle multiparton configurations on the Born level, 
but loop calculations are tough and lead to messy cancellations at
small resolution scales. Resummed matrix elements may circumvent
the latter problem for specific quantities, but then do not
provide exclusive accompanying events. Parton showers are based
on an improved leading-log (almost next-to-leading-log) approximation, 
and so cannot be accurate for well separated partons, but they offer a 
simple, process-independent machinery that gives a smooth blending of event 
classes (by Sudakov form factors) and a natural match to hadronization.
It is therefore natural to try to combine these descriptions, so
that ME results are recovered for widely separated partons while the
PS sets the subjet structure. 

For final-state showers in $Z^0 \to q\overline{q}$, such solu\-tions are 
the standard since long \cite{fsmatch}, e.g. by letting the shower 
slightly overpopulate the $q\overline{q}g$ phase space and then using
a Monte Carlo veto technique to reduce down to the ME level. This approach
easily carries over to showers in other colour-singlet resonance decays,
although the various relevant ME's have not all been implemented in
PYTHIA so far.

A similar technique is now available for the description of initial-state
radiation in the production of a single colour-singlet resonance, such
as $\gamma^*/Z^0/W^{\pm}$ \cite{gabriela}. The basic idea is to map the 
kinematics between the PS and ME descriptions, and to find a correction 
factor that can be applied to hard emissions in the shower so as to bring
agreement with the matrix-element expression. Some simple algebra
shows that, with the PYTHIA shower kinematics definitions,
the two $q\overline{q}' \to gW^{\pm}$ emission rates disagree by a 
factor
\[
R_{q\overline{q}' \to gW}(\hat{s},\hat{t}) = 
\frac{(\mathrm{d}\hat{\sigma}/\mathrm{d}\hat{t})_{\mathrm{ME}} }%
     {(\mathrm{d}\hat{\sigma}/\mathrm{d}\hat{t})_{\mathrm{PS}} } = 
\frac{\hat{t}^2+\hat{u}^2+2 m_W^2\hat{s}}{\hat{s}^2+m_W^4} ~, 
\]
which is always between $1/2$ and 1. 
The shower can therefore be improved in two ways, relative to the 
old description. Firstly, the maximum virtuality of emissions is 
raised from $Q^2_{\mathrm{max}} \approx m_W^2$ to 
$Q^2_{\mathrm{max}} = s$, i.e. the shower is allowed to populate the 
full phase space. Secondly, the emission rate for the final (which 
normally also is the hardest) $q \to qg$ emission on each side is 
corrected by the factor $R(\hat{s},\hat{t})$ above, so as to bring 
agreement with the matrix-element rate in the hard-emission region.
In the backwards evolution shower algorithm \cite{backwards}, this 
is the first branching considered.

The other possible ${\mathcal{O}}(\alpha_s)$ graph is $qg \to q'W^{\pm}$,
where the corresponding correction factor is
\[
R_{qg \to q'W}(\hat{s},\hat{t}) =
\frac{(\mathrm{d}\hat{\sigma}/\mathrm{d}\hat{t})_{\mathrm{ME}} }%
     {(\mathrm{d}\hat{\sigma}/\mathrm{d}\hat{t})_{\mathrm{PS}} } = 
\frac{\hat{s}^2 + \hat{u}^2 + 2 m_W^2 \hat{t}}{(\hat{s}-m_W^2)^2 + m_W^4} ~,
\]
which lies between 1 and 3. A probable reason for the lower shower 
rate here is that the shower does not explicitly simulate the $s$-channel 
graph $qg \to q^* \to q'W$. The $g \to q\overline{q}$ branching 
therefore has to be preweighted by a factor of 3 in the shower, but 
otherwise the method works the same as above. Obviously, the shower 
will mix the two alternative branchings, and the correction factor 
for a final branching is based on the current type.

The reweighting procedure prompts some other chang\-es in the shower. 
In particular, $\hat{u} < 0$ translates into a constraint on the phase
space of allowed branch\-ings. 

Our published comparisons with data on the $W$ $p_{\perp}$ spectrum 
show quite a good agreement with this improved simulation \cite{gabriela}. 
A worry was that an unexpectedly large primordial $k_{\perp}$, around 
4 GeV, was required to match the data in the low-$p_{\perp W}$ region. 
However, at that time we had not realized that the data were not fully 
unsmeared. The required primordial $k_{\perp}$ is therefore likely to
drop by about a factor of two \cite{joey}.

It should be noted that also other approaches to the same problem have
been studied recently. The HERWIG one requires separate treatments in 
the hard- and soft-emission regions \cite{herwigw}. Another, more 
advanced PYTHIA-based one \cite{steve}, also addresses the next-to-leading 
order corrections to the total $W$ cross section, while the one outlined 
above is entirely based on the leading-order total cross section. There 
is also the possibility of an extension to Higgs production \cite{steveh}, 
which is rather less trivial since already the leading-order cross section 
$gg \to H$ contains a QCD loop.

Summarizing, we now start to believe we can handle initial- and 
final-state showers, with next-to-leading-order accuracy, in cases where 
these can be separated by the production of colour singlet resonances
--- even if it should be realized that much work remains to cover the 
various possible cases. That still does not address the big class of 
QCD processes where the initial- and final-state radiation does not 
factorize. Possibly, correction factors to showers could be found
also here. Alternatively, it may become necessary to start showers from
given parton configurations of varying multiplicity and with
virtual-correction weights, as obtained from higher-order ME
calculations. So far, \mbox{PYTHIA} only implements a way to start from a
given four-parton topology in $e^+e^-$ annihilation, picking one
of the possible preceding shower histories as a way to set constraints
for the subsequent shower evolution \cite{johan}. This approach obviously 
needs to be extended in the future, to allow arbitrary parton 
configurations. Even more delicate will be the consistent treatment
of virtual corrections \cite{christerw}, where much work remains.

\section{Charm and Bottom Hadronization}

Significant asymmetries are observed between the production of
$D$ and $\overline{D}$ mesons in $\pi^- p$ collisions, with hadrons
that share some of the $\pi^-$ flavour content very much favoured at 
large $x_F$ in the $\pi^-$ fragmentation region \cite{casym}. This 
behaviour was qualitatively predicted by PYTHIA; in fact, the predictions
were for somewhat larger effects than seen in the data. The new data 
has allowed us to go back and take a critical look at the uncertainties 
that riddle the heavy-flavour description \cite{emanuel}. Many effects 
are involved, and we here constrain ourselves to only mentioning one.

A hadronic event is conventionally subdivided into sets of partons
that form separate colour singlets. These sets are represented by strings,
that e.g. stretch from a quark end via a number of intermediate gluons
to an antiquark end. Three string mass regions may be distinguished for 
the hadronization.
\begin{Enumerate}
\item {\em Normal string fragmentation}. In the ideal situation, each 
string has a large invariant mass. Then the standard iterative 
fragmentation scheme \cite{lund} works well. In practice, 
this approach can be used for all strings above some cut-off mass of a 
few GeV. 
\item {\em Cluster decay}.
If a string is produced with a \mbox{small} invariant mass, maybe only 
two-body final states are kinematically accessible. The traditional 
iterative Lund scheme is then not applicable. We call such a low-mass 
string a cluster, and consider it separately from above. In recent 
program versions, the modelling has now been improved to give a smooth 
match on to the standard string scheme in the high-cluster-mass limit.
\item {\em Cluster collapse}.
This is the extreme case of the above situation, where the string 
mass is so small that the cluster cannot decay into two hadrons.
It is then assumed to collapse directly into a sing\-le hadron, which
inherits the flavour content of the string endpoints. The original 
continuum of string/cluster masses is replaced by a discrete set
of hadron masses. Energy and momentum then cannot be conserved
inside the cluster, but must be exchanged with the local neighbourhood
of the cluster. This description has also been improved.  
\end{Enumerate}

In general, flavour asymmetries are predicted to be smaller for bottom 
than for charm, and smaller at higher energies (except possibly at very 
large rapidities). One can therefore not expect any spectacular 
manifestations at the Tevatron. However, other nontrivial features do 
not die out as fast, like a non-negligible systematic shift between the 
rapidity of a heavy quark and that of the hadron produced from it
\cite{emanuel}. The possibility of such effects should be considered 
whenever trying to extract any physics from heavy flavours.

\section{Multiple Interactions}

Multiple parton--parton interactions is the concept that, based on
the composite nature of hadrons, several parton pairs may interact
in a typical hadron--hadron collision \cite{maria}. Over the years, 
evidence for this mechanism has accumulated, such as the recent di\-rect
observation by CDF \cite{cdfmultint}. The occurences with two 
parton pairs at reasonably large $p_{\perp}$ just form the top of
the iceberg, however. In the PYTHIA model, most interactions are
at lower $p_{\perp}$, where they are not visible as separate jets but
only contribute to the underlying event structure. As such, they are
at the origin of a number of key features, like the broad 
multiplicity distributions, the significant forward--backward
multiplicity correlations, and the pedestal effect under jets.

Since the perturbative jet cross section is divergent for 
$p_{\perp} \to 0$, it is necessary to regularize it, e.g. by a
cut-off at some $p_{\perp\mathrm{min}}$ scale. That such a 
regularization should occur is clear from the fact that the incoming
hadrons are colour singlets --- unlike the coloured partons assumed in 
the divergent perturbative cal\-cu\-la\-tions --- and that therefore the 
colour charges should screen each other in the $p_{\perp} \to 0$ limit.
Also other damping mechanisms are possible \cite{goga}.
Fits to data typically give $p_{\perp\mathrm{min}} \approx 2$ GeV,
which then should be interpreted as the inverse of some colour
screening length in the hadron.   

One key question is the energy-dependence of $p_{\perp\mathrm{min}}$; 
this may be relevant e.g. for comparisons of jet rates at different 
Tevatron energies, and even more for any extrapolation to LHC energies. 
The problem actually is more pressing now than at the time of our 
original study \cite{maria}, since nowadays parton distributions are 
known to be rising more steeply at small $x$ than the flat $xf(x)$ 
behaviour normally assumed for small $Q^2$ before HERA. This 
translates into a more dramatic energy dependence of the 
multiple-interactions rate for a fixed $p_{\perp\mathrm{min}}$. 

The larger number of partons also should increase the amount of
screening, however, as confirmed by toy simulations \cite{johann}.
As a simple first approximation, $p_{\perp\mathrm{min}}$ is assumed
to increase in the same way as the total cross section, i.e. with some 
power $\epsilon \approx 0.08$ \cite{dl} that, via reggeon phenomenology,
should relate to the behaviour of parton distributions at small $x$ 
and $Q^2$. Thus the new default in PYTHIA is
\[
p_{\perp\mathrm{min}} = (1.9~{\mathrm{GeV}}) \left(
\frac{s}{1~\mathrm{TeV}^2} \right)^{0.08} ~.
\]

\section{Interconnection Effects}

The widths of the $W$, $Z$ and $t$ are all of the order of 
2 GeV. A Standard Model Higgs with a mass above 200 GeV, as well 
as many supersymmetric and other Beyond the Standard Model particles
would also have widths in the multi-GeV range. Not far from
threshold, the typical decay times 
$\tau = 1/\Gamma  \approx 0.1 \, {\mathrm{fm}} \ll  
\tau_{\mathrm{had}} \approx 1 \, \mathrm{fm}$.
Thus hadronic decay systems overlap, between a resonance and the
underlying event, or between pairs of resonances, so that the final 
state may not contain independent resonance decays.

So far, studies have mainly been performed in the context of
$W$ pair production at LEP2. Pragmatically, one may here distinguish 
three main eras for such interconnection:
\begin{Enumerate}
\item Perturbative: this is suppressed for gluon energies 
$\omega > \Gamma$ by propagator/time\-scale effects; thus only
soft gluons may contribute appreciably.
\item Nonperturbative in the hadroformation process:
normally modelled by a colour rearrangement between the partons 
produced in the two resonance decays and in the subsequent parton
showers.
\item Nonperturbative in the purely hadronic phase: best exemplified 
by Bose--Einstein effects.
\end{Enumerate}
The above topics are deeply related to the unsolved problems of 
strong interactions: confinement dyna\-mics, $1/N^2_{\mathrm{C}}$ 
effects, quantum mechanical interferences, etc. Thus they offer 
an opportunity to study the dynamics of unstable particles,
and new ways to probe confinement dynamics in space and 
time \cite{GPZ,ourrec}, {\em but} they also risk 
to limit or even spoil precision measurements.

A key gauge is the interconnection impact on $W$ mass measurements 
at LEP2. Perturbative effects are not likely
to give any significant contribution to the systematic error,
$\langle \delta m_W \rangle \lessim 5$~MeV \cite{ourrec}. 
Colour rearrangement is not understood from first principles,
but many models have been proposed to model effects 
\cite{ourrec,otherrec,HR}, and a conservative estimate gives 
$\langle \delta m_W \rangle \lessim 40$~MeV. 
For Bose--Einstein again there is a wide spread in models, and an 
even wider one in results, with about the same potential systematic
error as above \cite{ourBE,otherBE,HR}.
The total QCD interconnection error is thus below $m_{\pi}$ in 
absolute terms and 0.1\% in relative ones, a small number that 
becomes of interest only because we aim for high accuracy. 

A study of $e^+e^- \to t\overline{t} \to b W^+ \overline{b} W^- 
\to b \overline{b} \ell^+ \nu_{\ell} \ell'^- \overline{\nu}'_{\ell}$
near threshold gave a realistic interconnection 
uncertainty of the top mass of around 30 MeV, but also showed that 
slight mistreatments of the combined colour and showering structure 
could blow up this error by a factor of ten \cite{intertop}. 
For hadronic top decys, errors could be much larger.

The above numbers, when applied to hadronic physics, are maybe not
big enough to cause an immediate alarm. The addition of a coloured 
underlying event --- with a poorly-understood multiple-interaction
structure as outlined above --- has not at all been considered so far, 
however, and can only make matters worse in hadronic physics than in 
$e^+e^-$. This is clearly a topic for the future, where we
should be appropriately humble about our current understanding,
at least when it comes to performing precision measurements.

QCD interconnection may also be at the root of a number of 
other, more spectacular effects, such as rapidity gaps and the whole
Pomeron concept \cite{uppsalapom}, and the unexpectedly large rate of
quarkonium production \cite{uppsalaonia}.

\section{The Future: On To C++}

Finally, a word about the future. PYTHIA continues to be developed. 
On the physics side, there is a need to increase the support given 
to different physics scenarios, new and old, and many areas of the 
general QCD machinery for parton showers, underlying events and 
hadronization require further improvements, as we have seen. 

On the technical side, the main challenge is a transition from Fortran 
to C++, the language of choice for Run II (and LHC). To address this,
the PYTHIA 7 project was started in January 1998, with L. L\"onnblad as 
main responsible. A similar project, but more ambitious and better 
funded, is now starting up for HERWIG, with two dedicated postdoc-level 
positions and a three-year time frame. 

For PYTHIA, what exists today is a strategy document \cite{leif}, 
and code for the event record, the particle object, some particle data 
and other data base handling, and the event generation handler structure. 
All of this is completely new relative to the Fortran version, and is 
intended to allow for a much more general and flexible formulation of 
the event generation process. The first piece of physics, the string 
fragmentation scheme, is being implemented by M. Bertini, and is nearing 
completion. The subprocess generation method is being worked on for the 
simple case of $e^+e^- \to Z^0 \to q\overline{q}$. The hope is to have 
a ``proof of concept'' version soon, and some of the current 
PYTHIA functionality up and running by the end of 2000.
It will, however, take much further effort after that to provide a 
program that is both more and better than the current PYTHIA 6 
version. It is therefore unclear whether PYTHIA 7 will be of
much use during Run II, except as a valuable exercise for the future.

\end{document}